\documentclass{article}

\newcommand{\eref}[1]{Eq. (\ref{#1})}

\begin{document}
%\begin{frontmatter}
%\markboth{C. V. Usenko \and  B. I. Lev  }{Vacuum friction of uniformly accelerated detector}
	
\title{	Vacuum Friction of Uniformly Accelerated Detector}%

\author{C. V. Usenko\footnote{const.usenko@gmail.com}\quad  and\quad B. I. Lev\footnote{ bohdan.lev@gmail.com}}%
%\address{Bogolyubov Institute for Theoretical Physics\\
%	14-b Metrolohichna str. Kiev
%	Ukraine 03143
%}

%\author{}%

%\address{Bogolyubov Institute for Theoretical Physics\\
%	14-b Metrolohichna str. Kiev
%	Ukraine 03143}

\maketitle

	\begin{abstract}
	The analysis of uniformly accelerated longitudinally extended detector is performed and it is shown that the response of such a detector does not differ from the response of the Unruh detector, but the its excitation is caused not by the thermal bath, but by interaction with the fluctuations of the quantum field by virtual quanta.
	\end{abstract}

%\keywords{	Unruh effect, vacuum friction, uniform acceleration, unitary nonequivalent representations}

%\end{frontmatter}

%\linenumbers
	
\section{Introduction}
The Unruh effect \cite{Unruh} is a special case of one of the non-trivial mathematical properties of relativistic quantum field theory, namely the existence of unitary nonequivalent representations of a quantum field \cite{BD,EM,Grib}.
The nonequivalence of representations associated with inertial and uniformly accelerated reference frames is realized by the fact that the vacuum state of a quantum field registers by the uniformly accelerated detector as a state with noise distributed according to thermal law with a temperature proportional to the acceleration of the detector.

Since its discovery, the Unruh effect has been actively discussed in an application to a variety of problems.
Its effect on entanglement  between different observers \cite{DD,WJ11} as well as quantum non-locality \cite{TJ13,TWJ12} and quantum measurement techniques \cite{MS14,AF14} is investigated.  
Quantum information and quantum metrology \cite{TWJ15}, quantum steering \cite{steering} should also take into account the Unruh effect. 
 
 Recently, attention has been resumed to the study of the influence of quantum fluctuations on the non-uniformity  of charged particles motion \cite{Barnet:PRL,Barnet:JMO}.
This topic was studied from different perspectives but the results obtained did not yet find the final formulation. This is due to the fact that the problem statement must take into account two important aspects of quantum theory.

The first aspect is due to the influence of the non-uniformity  of motion of the particle on the vacuum fluctuations of the quantum field. It is well known that the representation of quantum fields in each inertial reference frame is unitary equivalent and have a common vacuum state. The accelerated systems have other representations that were actively investigated in \cite{BD, Grib} in order to analyze the physical content of the quantum evaporation of black holes. It turned out that the uniformly accelerating frame of reference has a representation of quantum fields, unitarily non-equivalent inertial, vacuum fluctuations can inhibit accelerated bodies (the Unruh effect \cite{Unruh}).

The second aspect concerns the causes of the non-uniformity  of the charged particle motion.  For charged particles, especially in the atoms (ions), the cause of the accelerated motion can be only the electromagnetic field. Nuclear or weak interaction acts on a very small scale in comparison with the uncertainty of the coordinates of atomic electrons, gravitational interaction due to the principle of equivalence must be characterized by the magnitude of the order of a square or cube of the ratio of the size of the atom to the radius of the curvature and is significant only in the vicinity of the singularity (under the horizon of black holes of stellar scales). The classical theory of the field of accelerated charges, begun in \cite{Born} and set forth in \cite{Ginzburg}, argues that it is necessary to take into account the delay of the moving charge field and to distinguish the near-field from the radiation zone. The results of investigations of the effects of vacuum fluctuations on a uniform charge motion are given in \cite{Thirring}. In addition to the classic representations of the feature of this effect is the need to take into account, in addition to delay, multi-photon radiation in a wide range of frequencies.
Subsequently, the Fulling-Davies-Unruh effect was discovered \cite{Myhrvold} in the quantum-field task about the motion of an electron in a uniform electric field.

We previously (1986) published a hard-to-reach preprint \cite{self} in which the problem of vacuum friction of a uniformly accelerated detector was investigated, which takes into account the general quantum approach to solving this problem.
 {Our results differ from \cite{Barnet:PRL, Barnet:JMO} in the absence of friction when the detector moves without acceleration. Subsequently, the absence of friction during inertial motion of the detector was repeatedly confirmed by other methods \cite{Iso,Parentani,Reznik}.}
 The work \cite{self} studied the physical energy source used for braking and showed that the energy to overcome vacuum friction comes from the force that provides acceleration. In the absence of acceleration, the frictional force is absent, as in classical theory. In this paper, we will retrieve the results obtained earlier for a better understanding of the problem and will indicate some way to solve it.

The interest in the question of the influence of the acceleration of the quantum particle detector on its operation is due to the difficulties that exist in the problem of determining single-particle states in a curved space-time. In different frames of reference in a curved space-time, single-particle states usually differ so significantly that the corresponding Fock representations are unitary nonequivalent \cite{Grib}. In some cases (static space-time or the presence of a sufficiently high symmetry group), there are physical prerequisites for the preferred choice of the frame of reference, but in general there are many physically equal reference frames, and with them a lot of different definitions of single-particle states. The elucidation of the physical content of the differences between these definitions is the main goal of the work on the study of a uniformly accelerated detector in flat space-time.

Rindler  in the paper \cite{RW} shows that in a uniformly accelerated frame of reference for a scalar massless field there exists a representation that exactly duplicates the representation of plane waves in an inertial reference frame. This representation  is unitarily nonequivalent to an inertial reference frame \cite{Unruh,BD}, the inertial vacuum looks as if (from the point of view of an uniformly accelerated observer) it was filled with particles distributed according to the usual law of thermal equilibrium with a temperature proportional to the acceleration of the reference frame. This result showed that the differences between the quantum field representations corresponding to different frames of reference are not the prerogative of the curved space-time.

The presence of nonequivalent representations in different frames of reference in a flat space-time makes it possible to study the physical content of these differences, since in one of the representations (corresponding to the inertial frame of reference) the physical content of the concepts of the single-particle and vacuum states is well known. Unruh in \cite{Unruh} investigated  the response of an uniformly  accelerated detector to a quantum field in the state of an inertial vacuum, a point monopole was used as a model of the detector, which, as a result of interaction with a quantum field, could become excited states. It was shown that the probability of detector excitation is distributed over the energy of the final states in the same way as if the detector were in contact with a quantum field heated to the predicted temperature by Rindler.

As a result of particle registration, the energy of the Unruh detector increases, even if the field whose particles the detector registers remains in the vacuum state. This emergence of energy "out of nowhere" was the basis for doubts about the correctness of the physical interpretation of Unruh's result, for example, in \cite{BD}. A few attempts were made to modify the formulation of the problem in order to obtain physically more obvious results \cite{AS}. However, the problem the sources of the energy absorbed by the detector have not been resolved. 

{ The state of the Unruh detector depends on the quantum field (internal degree of freedom) and the external field (spatial motion). It is generally accepted that the back action of the detector on the external field can be neglected. The question of the effect of the detector on a quantum field is not so simple. The simplest assumption that a detector can excite plane waves of a quantum field contradicts the causality principle; the field of detector is represented as a retarded Green's function. It would be more accurate to take into account the contribution of the virtual quanta of the field and the detector, similar to the dressing of an electron by a coat of virtual photons \cite{Myhrvold}. The renormalization of the charge that arises as a result of this account is accompanied by renormalization of the mass, which should affect the spatial motion of the detector \cite{Reznik}. We can eliminate the need for renormalization if we neglect the back action of the detector on the quantum field. Under this assumption, the energy of the quantum field is conserved.}

{In this paper, in order to identify the sources of energy wich absorbs by the detector, we consider a somewhat complicated model of the detector in which field-detector interaction  depends not on the internal energy of detector, but on distance between detector and uniformly accelerated leading center. As a result of this complication, it was found that the interaction of a uniformly accelerated detector with an inertial vacuum leads to the appearance of a force that inhibits the detector. This force is proportional to the acceleration of the leading center. To maintain the given acceleration of the detector this additional force must be compensated by an increase in the external force that causes the leading center to accelerate. Work done by an external force is the cause of energy release in the detector.}

\section{Uniformly accelerated Reference Frame}
The world line of a body moving with constant acceleration in two-dimensional space-time (flat) is described by equations
\begin{equation}\label{eq:one}
x^0=\frac{1}{a}\sinh{at},\quad x^1=\frac{1}{a}\cosh{at},
\end{equation}
here $a$ is acceleration, $t$ is proper time. 

In contrast to the inertial motion, the hyper-surfaces of constant time (orthogonal to the world line) do not remain parallel, but constantly rotate, crossing at the same point, as shown in \cite{Grib} . For this point, it is impossible to continue the hypersurfaces of constant time in an uniformly accelerated frame of reference, because it covers only a part of space-time. This part is one quadrant bounded by the horizons of the past $ H_- $ and the future $ H_+ $. These horizons are also hypersurfaces of constant time, but related to infinitely distant moments of proper time. Coordinates in \eref{eq:one} are chosen so that the point of intersection of the horizons corresponds to the coordinates $x^0=0$, $x^1=0$.

The spatial coordinate $ y $ in the uniformly accelerated frame of reference is related to the Cartesian coordinates by the relation
\begin{equation}\label{eq:two}
y= \sqrt{{x^0}^2-{x^1}^2}-\frac{1}{a}
\end{equation}
 and represents (measured along the hypersurface of a constant time) the interval between the point under consideration and the world line of the body. In the coordinates $ y $, $ t $, the space-time metric has the form
\begin{equation}\label{eq:3}
ds^2=\left(1+ay\right)^2dt^2-dy^2
\end{equation}
and is singular on the horizons, which is additional evidence of the limitedness of that part of the space-time that can be covered by the uniformly accelerated frame of reference.

The metric \eref{eq:3} is conformally flat, as is easily seen after replacing the spatial coordinate by
  \begin{equation}\label{eq:4}
  \xi=\frac{1}{a}\ln\left(1+ay\right).
  \end{equation}

In the future, these coordinates will not be used, but it is useful to note that it is in coordinates that a scalar massless field can be represented in the form of an expansion in "plane waves", as in an inertial reference frame. For the future, it is more important that the metric \eref{eq:3} has a Killing vector 
$ \frac{\partial}{\partial t} $, corresponding to the symmetry with respect to translations along time.
This means that in the uniformly accelerated reference system there is a law of conservation of energy, as in the inertial reference system, with the difference that the energy of the uniformly accelerated frame of reference in Cartesian coordinates corresponds not to the Killing vector $ \frac{\partial} {\partial x^0} $ that generates energy in the inertial reference frame, but the Killing vector of Lorentz rotations 
$ x^1 \frac{\partial} {\partial x^0} + x^0 \frac{\partial} {\partial x^1} $.
This means that the concept of energy in a uniformly accelerated frame of reference is fundamentally different from the energy of the inertial frame of reference (the corresponding operators do not commute).

  The world line \eref{eq:one} in the coordinates $ y $, $ t $ is the straight line $ y = 0 $. Further, it will be assumed that the quantum field detector moves along this straight  line.

\section{Quantum field Representations}
The usual representation of a massless scalar field in two-dimensional space-time in the form of a superposition of plane waves
%\begin{widetext}
\begin{equation}\label{eq:5}
\phi\left(x^0,x^1 \right) =
 \int\limits_{-\infty}^{\infty}
{\frac{dk}{\sqrt{2\pi \omega}}\left\lbrace e^{i\omega x^0 - i k x^1} a_k+ e^{-i\omega x^0 + i k x^1} a_k^+ \right\rbrace },\quad
\omega=\left| k\right|,
\end{equation} 
can be conveniently modified somewhat by selecting specially waves to the right and to the left
\begin{equation}\label{eq:6}
\begin{array}{r}
\phi\left(x^0,x^1 \right) = %\\
\int\limits_{0}^{\infty}
\frac{d\omega}{\sqrt{2\pi \omega}}\left\lbrace e^{i\omega\left(x^0 -  x^1 \right)  } \overrightarrow{a}_{\omega}+ e^{-i\omega \left(x^0 -  x^1 \right)} \overrightarrow{a}_{\omega}^+ \right\rbrace+\\
\int\limits_{0}^{\infty}\frac{d\omega}{\sqrt{2\pi \omega}}\left\lbrace e^{i\omega\left(x^0 +  x^1 \right)  } \overleftarrow{a}_{\omega}+ e^{-i\omega \left(x^0 +  x^1 \right)} \overleftarrow{a}_{\omega}^+	\right\rbrace .
\end{array}
	\end{equation}
in the coordinates $ t $, $ y $ this representation has the form
  \begin{equation}\label{eq:7}
\begin{array}{r}
 \phi\left(t,y\right) =  %\\
\int\limits_{0}^{\infty}
\frac{d\omega}{\sqrt{2\pi \omega}} 
\left\lbrace 
e^{-i\omega\left(a^{-1}+y \right)e^{-at}  } \overrightarrow{a}_{\omega} + 
e^{i\omega\left(a^{-1}+y \right)e^{-at}} \overrightarrow{a}_{\omega}^+ 
\right\rbrace+\\
\int\limits_{0}^{\infty}\frac{d\omega}{\sqrt{2\pi \omega}}
\left\lbrace e^{i\omega\left(a^{-1}+y \right)e^{at} } \overleftarrow{a}_{\omega}+ e^{-i\omega\left(a^{-1}+y \right)e^{at}} \overleftarrow{a}_{\omega}^+	\right\rbrace .
\end{array} 
 \end{equation}
Partial waves 
\begin{equation}\label{eq:8}
\overrightarrow{\phi}^{\left(\pm \right) }_{\omega}\left(x^0, x^1 \right)=\frac{e^{\pm i\omega\left(x^0 -  x^1 \right)  }}{\sqrt{2\pi \omega}};\quad
\overleftarrow{\phi}^{\left(\pm \right) }_{\omega}\left(x^0, x^1 \right)=\frac{e^{\pm i\omega\left(x^0 +  x^1 \right)  }}{\sqrt{2\pi \omega}},
\end{equation}
satisfy the orthogonality condition 
\begin{equation}\label{eq:9}
\int\limits_{-\infty}^{\infty}
\left[
\phi_{1}\left(x^0, x^1 \right)\frac{\partial}{\partial x^0}\phi_{2}\left(x^0, x^1 \right)- 
 \frac{\partial}{\partial x^0}\phi_{1}\left(x^0, x^1 \right)\phi_{2}\left(x^0, x^1 \right)
 \right] \frac{i\ dx^1}{2}
 =\delta_{1,2},
\end{equation}
%\end{widetext} 
when integrating over the entire space, which corresponds to the use of the hyper-surfaces of the time constant of the inertial reference frame; in this connection, the representations \eref{eq:5} - \eref{eq:7} correspond to the inertial frame, and the state of the quantum field, annihilating by all the annihilation operators $ \overrightarrow{a}{\omega} $, $ \overleftarrow{a}{\omega} $, is an inertial vacuum.

By symmetry of the metric \eref{eq:3} with respect to the Killing vector, we can construct another representation \cite{RW}, in which
%\begin{widetext}
\begin{equation}\label{eq:10}
\begin{array}{r}
\phi\left(t,y\right) = %\\
 \int\limits_{0}^{\infty}
\left\lbrace 
\left(1+ay \right)^{-i\nu}e^{i\nu t} \overrightarrow{b}_{\nu} + 
\left(1+ay \right)^{i\nu}e^{-i\nu t} \overrightarrow{b}_{\nu}^+ 
\right\rbrace \frac{d\nu}{\sqrt{2\pi \nu}}+\\
\int\limits_{0}^{\infty}
\left\lbrace \left(1+ay \right)^{i\nu}e^{i\nu t} \overleftarrow{b}_{\nu}+ 
\left(1+ay \right)^{-i\nu}e^{-i\nu t} \overleftarrow{b}_{\nu}^+	\right\rbrace 
\frac{d\nu}{\sqrt{2\pi \nu}}.
\end{array}
\end{equation}
Partial waves for this representation 
\begin{equation}\label{eq:11}
\overrightarrow{\phi}^{\left( \pm\right) }_{\nu}\left(t,y\right)=
\frac{\left(1+ay \right)^{\mp i\nu}e^{\pm i\nu t}}{\sqrt{2\pi \nu}};\quad
\overleftarrow{\phi}^{\left( \pm\right) }_{\nu}\left(t,y\right)=
\frac{\left(1+ay \right)^{\pm i\nu}e^{\pm i\nu t}}{\sqrt{2\pi \nu}},
\end{equation}
are orthogonal only on constant-time hypersurfaces of the uniformly accelerated reference frame
\begin{equation}\label{eq:12}
\frac{i}{2}\int\limits_{-a^{-1}}^{\infty}
\left(
\phi_{1}\left(t, y\right)\frac{\partial\phi_{2}\left(t, y \right)}{\partial t}-
\frac{\partial\phi_{1}\left(t, y \right)}{\partial t}\phi_{2}\left(t, y \right)
\right) \frac{dy}{y+a^{-1}}
=\delta_{1,2},
\end{equation}
%\end{widetext}
so that the representation corresponds to an uniformly accelerated frame of reference.

The partial waves \eref{eq:11}, and with them the representation of the quantum field \eref{eq:10},
 are not defined on the whole space-time, but only on the quadrant belonging to the uniformly accelerated frame. 
 Therefore, the operators $ b $, $ b ^ + $ realize the representation not of the entire algebra of the observable quantum field, 
 but only of its subalgebra having a support on this quadrant. 
 In accordance with the general results of the theory of representations of algebras of observables, 
 in this case the representation is not exact \cite{EM}, and, accordingly, the representation \eref{eq:10} is 
 unitarily non-equivalent \eref{eq:5}. To see this, 
 we consider the Bogolyubov transformations connecting the  representations \eref{eq:6} and \eref{eq:10}:
\begin{eqnarray}\label{eq:13}
\overrightarrow{b}_{\nu}=\int\limits_0^{\infty}d\omega
 \left\lbrace 
\overrightarrow{\alpha}_{\nu}\left(\omega \right) \overrightarrow{a}_{\omega}+
\overrightarrow{\beta}_{\nu}\left(\omega \right) \overrightarrow{a}_{\omega}^+
\right\rbrace; \\ \label{eq:14}
\overleftarrow{b}_{\nu}=\int\limits_0^{\infty}d\omega
\left\lbrace 
\overleftarrow{\alpha}_{\nu}\left(\omega \right) \overleftarrow{a}_{\omega}+
\overleftarrow{\beta}_{\nu}\left(\omega \right) \overleftarrow{a}_{\omega}^+
\right\rbrace, 
\end{eqnarray}
The Bogolyubov coefficients are
\begin{eqnarray}\label{eq:15}
\overrightarrow{\beta}_{\nu}\left( \omega\right) =\frac{\sqrt{a\nu}}{\pi}
\left(\frac{a}{\omega} \right)^{i\nu+\frac{1}{2}}
e^{\frac{\pi \nu}{2}} 
\Gamma\left(i\nu \right) 
; \quad \overrightarrow{\alpha}_{\nu}\left( \omega\right) =\sqrt{1+\left|\overrightarrow{\beta}_{\nu}\left( \omega\right) \right|^2}
\\ \label{eq:16}
\overleftarrow{\beta}_{\nu}\left( \omega\right) =-\frac{\sqrt{a\nu}}{\pi}
\left(\frac{a}{\omega} \right)^{-i\nu+\frac{1}{2}}
e^{\frac{\pi \nu}{2}} 
\Gamma\left(-i\nu \right) ;
\quad \overleftarrow{\alpha}_{\nu}\left( \omega\right) =\sqrt{1+\left|\overleftarrow{\beta}_{\nu}\left( \omega\right) \right|^2}
, 
\end{eqnarray}
for them the boundedness condition
\begin{equation}\label{eq:17}
\int\limits_0^{\infty}d\nu\int\limits_0^{\infty}d\omega\left|{\beta}_{\nu}\left( \omega\right) \right|^2 \le \infty 
\end{equation}
is not satisfied, which proves the unitary nonequivalence of the representations.

Of particular interest is the distribution of Rindler particles in a state that will be vacuum in an inertial frame of reference. To calculate this distribution, we consider the average over the inertial vacuum from the product of the creation and annihilation operators of Rindler particles
\begin{equation}\label{eq:18}
\left\langle  \overrightarrow{b}_{\nu'}^+\overrightarrow{b}_{\nu}\right\rangle =
\int\limits_0^{\infty}\overrightarrow{\beta}_{\nu'}^*\left( \omega\right)\overrightarrow{\beta}_{\nu}\left( \omega\right) d \omega.
\end{equation}
After the substitution \eref{eq:15}, the integral over $ \omega $ is calculated by changing the variables $ z = \ln \frac{\omega}{a} $ (it reduces to $ \delta $ -function), which gives
\begin{equation}\label{eq:19}
\left\langle  \overrightarrow{b}_{\nu'}^+\overrightarrow{b}_{\nu}\right\rangle =
\frac{4a^2}{e^{2\pi\nu}-1}\delta\left(\nu-\nu' \right). 
\end{equation}
Replacing the parameter $\nu$ by the Rindler frequency $ \omega = \nu a $, we obtain for the probability density of the registration of Rindler particles in the state of an inertial vacuum the following expression \cite{RW}:
\begin{equation}\label{eq:20}
\overrightarrow{n}\left(\omega \right) =\frac{1}{e^{\frac{2\pi}{a}\omega}-1},
\end{equation}
 equivalent to the thermal distribution with temperature $ T = \frac{a}{2 \pi} $. Similar expressions are obtained for waves traveling to the left.
 
 { The inertial vacuum state is similar but not equal to the  thermal bath state. Vacuum is a pure state with zero entropy, while   thermal bath is a mixture of pure states with entropy proportional to temperature. The entanglement investigated in [6,12] is a property of virtual quanta in vacuum state, not a thermal bath state. It is maximal for pure states and is absent in the state of a thermal bath.}

Since the representation of the quantum field in the uniformly accelerated frame of reference \eref{eq:10} is not defined on the whole space-time, the inverses to \eref{eq:13}, \eref{eq:14} do not exist - the state of the quantum field in one a uniformly accelerated frame of reference is not sufficient for a complete description of the quantum field in an inertial frame of reference. Therefore, the formulation of the problem of the distribution of inertial particles in the state of the Rindler vacuum is incorrect.

The existence in the plane space-time of quantum field representations, in which the definitions of single-particle states (and vacuum) differ so significantly, leads to the need to study the physical content of the differences in these representations. For this purpose, the work of the uniformly accelerated detector is analyzed.

\section{Uniformly accelerated Detector}

The detector model proposed by  Unruh \cite{Unruh} is a material dot with an internal structure. In this somewhat contradictory definition we mean that the spatial sizes of the detector in the Unruh model are assumed to be negligibly small, but there are, nevertheless,  discrete levels $ E_n $ of internal energy. 
{The interaction of the detector with a quantum field is determined by the Hamiltonian}
\begin{equation}\label{eq:21}
H_u=\hat{m}\phi(x^0\left( t\right) ,x^1\left( t\right)),
\end{equation}
containing the value of the quantum field on the world line of the detector $ \left\lbrace x^0\left( t\right) ,x^1\left( t\right) \right\rbrace  $ only and operator of the monopole moment $ \hat{m}  $ of the detector.

 Attention to such a model is due to the fact that in the space-time with arbitrary curvature, the existence of stationary states of spatially extended bodies is practically impossible, since even for classical bodies severe restrictions are imposed on the rigid motion. For an uniformly-accelerated detector, the use of the Unruh model is not necessary, since in an uniformly accelerated frame of reference there exists a symmetry with respect to shifts along the proper time.

As a more realistic model of the detector, it is useful to consider a system consisting of a leading center - a classical body moving along a given world line and a test particle - a quantum particle potentially connected with a leading center and interacting with a detectable quantum field.

The presence of the Killing vector $ \frac{\partial} {\partial t} $ allows for the uniformly accelerated motion of the leading center to set the interaction potential of the test particle with the leading center such that the test particle has stationary states. For this it is sufficient that in the coordinates $ t $, $ y $ associated with the center, the potential energy of the test particle depends only on the deviation $ y $ of the test particle from the world line of the leading center and does not depend on the intrinsic time $ t $.

{This assumption allows one to write the Hamiltonian of the test particle in the field of the leading  center as the sum of the kinetic and potential energy operators.}
\[\hat{H}=\frac{p^2}{2m} +U(y).\]

{Here it is taken into account that in the frame of reference accompanying the leading center the motion of the test particle remains non-relativistic.}
 
Suppose now that the interaction Hamiltonian of a test particle with a quantum field also depends on the displacement
$ y $
 of the test particle relative to the leading center. Then, in the first approximation, this Hamiltonian 
of interaction
can be written in the form
\begin{equation}\label{eq:22}
H_I=-qy\phi(t,y).
\end{equation}
where $ q $ is the "charge" of the test particle.

The interaction of the test particle of the detector with the quantum field leads to the fact that the energy of the test particle changes - increases with absorption of the field quanta and decreases with their emission. If the probe particle were left to itself, then in the 'test particle + quantum field' system, the total energy was conserved and a dynamic equilibrium would be established between the detector and the quantum field. However, the interaction of the test particle with the classical leading center should lead to the transfer of the energy absorbed from the field to the leading center. Due to the classical nature, the leading center is a sufficiently large system in which the energy obtained from the test particle dissipates. Thus, the interaction of a test particle with a quantum field can be registered by the leading center as the transfer of energy to the leading center.

 The characteristic of the operation of the detector is obviously the amount of energy absorbed by the test particle from the field per unit time, taking into account the fact that the state of the test particle as a result of interaction with both the quantum field and the leading center does not change. This energy absorbed per unit time can be calculated in a standard way, as an average value
\begin{equation}\label{eq:23}
\dot{Q}=-q\overline{y\frac{\partial}{\partial t}\phi(t,y)}.
\end{equation}
In this expression, the bar denotes averaging over both the state of the quantum field and the state of the test particle.

It should be noted that the state of the test particle is far from necessarily the ground one. If the leading center is not at zero temperature, the interaction of the leading center with the test particle should lead to the fact that the state of the test particle will be thermal with the same temperature. Therefore, it would be careless to use for its averaging over the state of the test particle its wave function of the ground state.

\section{Test particle Response}
To calculate the energy absorbed by the detector, it is possible to assume without attempting to refine the characteristics of the probe particle of the detector and its interaction with the leading center, that under the action of the external field the test particle acquires a  shift $ y \left(t \right) $, which in the first approximation is proportional to the field, in accordance with the methods of the theory of linear response, this assumption allows us to write the following expression for the displacement:
\begin{equation}\label{eq:24}
y\left(t\right)=y_f+\int\limits_0^{\infty}d\tau\alpha\left(\tau\right)\phi\left(t-\tau,0\right),
\end{equation}
in which {$ y_f $ is the free part of shift operator,} $ \alpha \left(\tau \right) $ is the response function of the detector, and the values of the quantum field $\phi\left(t-\tau,0\right)$ are computed on the world line of the leading center, since taking into account the displacement of $ y $ in the quantum field argument in this expression would be an excess of accuracy.

In this approximation, for the energy absorbed by the detector per unit time, we obtain
\begin{equation}\label{eq:25}
\dot{Q}\left(t \right) =-q\int\limits_0^{\infty}d\tau\alpha\left(\tau\right)
\left\langle \phi\left(t-\tau,0\right)\frac{\partial}{\partial t}\phi\left(t,0\right)\right\rangle 
\end{equation}
where the angle brackets mean that the average must be calculated only by the state of the quantum field.

Since the test particle is displaced from the equilibrium position by the action of the quantum field, 
a restoring force must act on the side of the leading center, which obviously compensates for the biasing force with which the 
quantum field acts on the test particle. To calculate this force, it is sufficient to calculate the average value of the 
commutator $ i \left[p, \ H_I \right] $. With \eref{eq:24} for this average strength, we get the following expression:
\begin{equation}\label{eq:26}
F\left(t\right)=-2q\int\limits_0^{\infty}{d\tau\alpha\left(\tau\right)
\left\langle \phi\left(t-\tau,0\right)\frac{\partial}{\partial y}\phi\left(t,0\right)\right\rangle }
\end{equation}
Let us now consider the response of the detector to a quantum field in the Rindler vacuum state. 
For this, it is necessary to use the representation of the quantum field in the form \eref{eq:10} and take into account 
that in the Rindler vacuum state the mean values of the Fock operators $ \overrightarrow{b}_\nu $, $ \overleftarrow{b}_\nu $ are equal to
\begin{equation}\label{eq:27}
	\left\langle \overrightarrow{b}_\nu \overrightarrow{b}^+_{\nu'}\right\rangle =
	\left\langle \overleftarrow{b}_\nu \overleftarrow{b}^+_{\nu'}\right\rangle =
	\delta_{\nu,\nu'},
	\end{equation}
and the averages of the remaining bilinear forms vanish. Direct calculation gives
\begin{equation}\label{eq:28}
\dot{Q}=-q\int\limits_0^{\infty}d\tau\alpha\left(\tau\right)\int\limits_0^{\infty}\frac{d\nu}{2\pi\nu}
\left(-2i\nu a\right)e^{i\nu a\tau}
\end{equation}
The integral over $ \nu $ in this expression is calculated using the usual substitution $ \nu \mapsto \nu + i0 $, which allows obtaining the final expression
\begin{equation}\label{eq:29}
\dot{Q}=\frac{q}{\pi}\int\limits_0^{\infty}\frac{d\tau}{\tau}\alpha\left(\tau\right),
\end{equation}
coincides with the expression for the response of an inertial detector to inertial vacuum.
This value is infinite if the response is switched on abruptly.
($\alpha\left(0 \right)\neq 0 $). 
It can be finite only with a gradual increase in the response of the detector.
($\alpha\left(\tau \right)\propto \tau^q,\quad \left| q\right| >0 $).
In other words, the detector response should lag behind.

A similar calculation for the mean force leads to expression
\begin{equation}\label{eq:30}
F\left(t\right)=-2q\int\limits_0^{\infty}d\tau\alpha\left(\tau\right)\int\limits_0^{\infty}\frac{d\nu}{2\pi\nu}
\left(i\nu ae^{i\nu a\tau} -i\nu ae^{i\nu a\tau}\right),
\end{equation}
which vanishes identically.

Thus, from the point of view of an uniformity  accelerated detector, a quantum field in the state of an uniformly accelerated vacuum behaves in the same way as a field in the state of an inertial vacuum from the point of view of an inertial detector.

The response of a uniformly accelerated detector to a quantum field in the state of an inertial vacuum, naturally, differs from \eref{eq:29}, \eref{eq:30}. To calculate this response, we use the fact that in the state of an inertial vacuum, analogous \eref{eq:27} relations are satisfied for the operators $ \overrightarrow{a}_\omega $, $ \overleftarrow{a}_\omega $. Taking this into account, the expression for the average energy absorbed by the detector per unit time has the form

\begin{equation}\label{eq:31}
\dot{Q}=-\frac{iq}{2\pi}\int\limits_0^{\infty}d\tau\alpha\left(\tau\right)\int\limits_0^{\infty}d\omega
% K_Q
\left( 
 e^{-at}e^{i\omega a^{-1}e^{-at}\left(e^{a\tau}-1 \right)}+
 e^{at}e^{i\omega a^{-1}e^{at}\left(1-e^{-a\tau} \right)}
 \right)
\end{equation}
%\[ K_Q=\left( 
%e^{-at}e^{i\omega a^{-1}e^{-at}\left(e^{a\tau}-1 \right)}+
%e^{at}e^{i\omega a^{-1}e^{at}\left(1-e^{-a\tau} \right)}
%\right) \]
The calculation of the integral over $ \omega $ is carried out, as usual, and gives
\begin{equation}\label{eq:32}
\dot{Q}=\frac{qa}{2\pi}\int\limits_0^{\infty}d\tau\alpha\left(\tau\right)
\frac{e^{a\tau}+1}{e^{a\tau}-1}.
\end{equation}
It is useful to note that in the limit $ a \mapsto 0 $ this expression goes to \eref{eq:29}. 
For comparison \eref{eq:32} with the results of Unruh \cite{RW} we consider the Fourier representation of the response function
\begin{equation}\label{eq:33}
\alpha\left(\tau\right)=\frac{1}{2\pi}\int\limits_{-\infty}^{\infty}d\omega\alpha\left(\omega\right)
e^{-i\omega\tau}
\end{equation}
Replacing the variable $ x = e^{- a \tau} $ 
yields the following expression:
%\[ \dot{Q}= \]
\begin{equation}\label{eq:34}
\dot{Q}=\frac{q}{2\pi^2}\int\limits_{-\infty}^{\infty}d\omega\alpha\left(\omega\right)\left[ 
\frac{a}{i\omega}+
\sum\limits_{k=1}^{\infty}\frac{2}{k+i\omega/a}
\right] 
\end{equation}
 Taking into account the analyticity of the response function in the upper half-plane, the dispersion relation
\[
\alpha\left(i p\right)=\frac{2}{\pi}
\int\limits_{0}^{\infty}\frac{\omega d\omega}{p^2+\omega^2}\alpha"\left(\omega\right),
\]
and decomposition
\[
\coth\left(\omega\right)=\frac{1}{\pi\omega}+
\frac{2}{\pi}\sum\limits_{k=1}^{\infty}\frac{\omega }{k^2+\omega^2},
\]
we obtain
\begin{equation}\label{eq:38}
\dot{Q}=\frac{q}{\pi}\int\limits_{0}^{\infty}d\omega\alpha"\left(\omega\right)
\coth\left(2\frac{%\pi
	\omega}{a}\right)
\end{equation}
This expression, like the Callen-Welton formula, contains only the imaginary part $ \alpha"\left(\omega \right) $ of the response function. Thus, the response to the inertial state of a quantum field in an uniformly accelerated detector is similar to the response of an inertial detector to a field in a thermodynamically equilibrium state with a temperature $ T = {a/2} $ in full accordance with Unruh's result.

The average value of the force acting on the detector from the  quantum field is calculated similarly. From \eref{eq:26} we have
\begin{equation}\label{eq:39}
F=2q\int\limits_0^{\infty}\frac{d\omega}{2\pi\omega}\int\limits_0^{\infty}d\tau\alpha\left(\tau\right)
%K_F
 \left[
 i\omega e^{-at}e^{i\omega a^{-1}e^{-at}\left(e^{a\tau}-1 \right)}-i\omega 
 e^{at}e^{i\omega a^{-1}e^{at}\left(1-e^{-a\tau} \right)}
 \right]
\end{equation}
%\[ K_F= \left[
%i\omega e^{-at}e^{i\omega a^{-1}e^{-at}\left(e^{a\tau}-1 \right)}-i\omega 
%e^{at}e^{i\omega a^{-1}e^{at}\left(1-e^{-a\tau} \right)}
%\right]\]
and after calculating the integral over $ \omega $ we obtain
\begin{equation}\label{eq:40}
F=\frac{aq}{\pi}\int\limits_0^{\infty}d\tau\alpha\left(\tau\right)
\left(\frac{1}{e^{a\tau}-1}- \frac{1}{1-e^{-a\tau}}\right)
\end{equation}
Introducing the notation 
\begin{equation}\label{eq:41}
\alpha_{l\tau}=\int\limits_0^{\infty}d\tau\alpha\left(\tau\right)
\end{equation}
for the response to a static perturbation, we finally obtain
\begin{equation}\label{eq:42}
F=-\frac{q\alpha_{l\tau}}{\pi}a
\end{equation}
These expressions determine the response of an uniformly accelerated detector to the vacuum (inertial) state of the quantum field.
\section{Conclusions}
Discussion of the results it is useful to begin the question of the source of energy absorbed by an uniformly accelerated detector when interacting with a quantum field in the state of an inertial vacuum. It should be noted that
 the energy of an uniformly accelerated detector is not conserved. The acceleration with which the detector moves is evidently due to the action of an external force, which constantly does work to move the detector. The interaction of the test particle with the leading center of the detector leads to the fact that a force acts on the test particle, which imparts the necessary acceleration to the test particle.

The action of quantum field in the state of an inertial vacuum on the detector leads to the appearance of a force F \eref{eq:42}, which prevents the motion of the test particle. Therefore, to maintain the given acceleration of this particle, the leading center must act with an additional force compensating F. This additional force obviously performs additional work, which is responsible for the energy absorption by the detector. Thus, it can be asserted that the energy absorption by the detector is not due to the energy of the quantum field but does by the work of accelerating forces.

Here we see an analogy with the release of heat in a conductor moving in a magnetic field.

The force F \eref{eq:42} acting on the detector from the side of the quantum field in the state of inertial vacuum, opposes the detector acceleration regardless of the direction of its motion and in this sense is similar to the frictional force. The reason for the action of this force is, as follows from \eref{eq:26}, the existence of correlations between the quantum field and the displacement of the detector. Such correlations exist already at the level of vacuum fluctuations of the quantum field, which are usually interpreted in terms of virtual quanta. In these terms, the expressions \eref{eq:25} and \eref{eq:26} for the absorbed energy and force correspond to the change in the energy and momentum of the detector as a result of the virtual quantum absorption at instant $ t=0 $ and its subsequent emission at current instant $ t $. At the same time, due to the dissipation of the excitation of the probe particle of the detector in its leading center, a part of the energy and momentum is absorbed, therefore, even with the inertial motion of the detector, the energy absorption rate is nonzero. In inertial motion, the rate of absorption of the pulse is also different from zero, but the rate of absorption of the pulse from the waves arriving from the right and from the left is the same, so the average force acting on the detector vanishes. If the detector moves non-inertially, the contributions of virtual particles are obviously not the same, the average force being greater on the side to which the detector is moving.

It should be noted significance of stationarity of the test particle states in the case of the uniformly accelerated detector, This stationarity produces the possibility of the existence of a stationary state of a test particle in which the leading center dissipates all  energy absorbed by the test particle.
In the curved space-time, in the general case, stationary states of the test particle are impossible, therefore, the transfer of the results given above to the detector in curved space-time requires additional studies.

\section*{Acknowledgement}
Authors want to thank to  S.P. Lukyanets, that he drew our attention to the importance of the results obtained by us on the vacuum friction of the uniformly accelerated detector.

%\section*{References}
\bibliographystyle{unsrt}

\bibliography{./vf}

\end{document}